\documentclass[twocolumn,tighten]{aastex62}

\usepackage{graphicx}
\usepackage{epstopdf}
\usepackage{hyperref}

\received{June 27, 2018}
\revised{}
\submitjournal{ApJL}

\shorttitle{Red vs. Blue: Early SNe~Ia color evolution}
\shortauthors{Stritzinger et al.}

\begin{document}

\title{Red vs Blue: Early observations of thermonuclear supernovae reveal two distinct populations?}

\correspondingauthor{M. Stritzinger}
\email{max@phys.au.dk}

\author[0000-0002-5571-1833]{Maximilian D. Stritzinger}
\affil{Department of Physics and Astronomy, Aarhus University, Ny Munkegade 120, DK-8000 Aarhus C, Denmark}
\affiliation{Visiting Astronomer, Institute for Astronomy, University of Hawai'i, 2680 Woodlawn Drive, Honolulu, HI 96822, USA}

\author{Benjamin J. Shappee}
\affiliation{Institute for Astronomy, University of Hawai'i, 2680 Woodlawn Drive, Honolulu, HI 96822, USA}

\author{Anthony L. Piro}
\affiliation{The Observatories of the Carnegie Institution for Science, 813 Santa Barbara St., Pasadena, CA 91101, USA}

\author{Christopher Ashall}
\affiliation{Department of Physics, Florida State University, 77 Chieftain Way, Tallahassee, FL, 32306, USA}

\author{E. Baron}
\affiliation{Homer L. Dodge Department of Physics and Astronomy, University of Oklahoma, 440 W. Brooks, Rm 100, Norman, OK 73019-2061, USA}
\affiliation{Visiting Astronomer, Hamburger Sternwarte, Gojenbergsweg 112, 21029 Hamburg, Germany}

\author{Peter Hoeflich}
\affiliation{Department of Physics, Florida State University, 77 Chieftain Way, Tallahassee, FL, 32306, USA}

\author{Simon Holmbo}
\affil{Department of Physics and Astronomy, Aarhus University, Ny Munkegade 120, DK-8000 Aarhus C, Denmark}

\author{Thomas W.-S. Holoien}
\affiliation{The Observatories of the Carnegie Institution for Science, 813 Santa Barbara St., Pasadena, CA 91101, USA}

\author{M. M. Phillips}
\affiliation{Carnegie Observatories, Las Campanas Observatory, Casilla 601, La Serena, Chile}

\author{C. R. Burns}
\affiliation{The Observatories of the Carnegie Institution for Science, 813 Santa Barbara St., Pasadena, CA 91101, USA}

\author{Carlos Contreras}
\affiliation{Carnegie Observatories, Las Campanas Observatory, Casilla 601, La Serena, Chile}

\author{Nidia Morrell}
\affiliation{Carnegie Observatories, Las Campanas Observatory, Casilla 601, La Serena, Chile}

\author{Michael A. Tucker}
\affiliation{Institute for Astronomy, University of Hawai'i, 2680 Woodlawn Drive, Honolulu, HI 96822, USA}

\begin{abstract}

We examine the early phase intrinsic $(B-V)_{0}$  color evolution of a dozen Type~Ia supernovae discovered within three days of the inferred time of first light ($t_{first}$)  and have  $(B-V)_0$ color information beginning within 5 days of $t_{first}$. The sample indicates there are two distinct early populations.  The first is a population exhibiting blue colors that slowly evolve, and the second population exhibits red colors and evolves more rapidly.  We find that the early-blue events are all 1991T/1999aa-like with more luminous slower declining light curves than those exhibiting early-red colors. Placing the first sample on the Branch diagram (i.e., ratio of \ion{Si}{2} $\lambda\lambda$5972, 6355 pseudo-Equivalent widths) indicates all blue objects are of the Branch Shallow Silicon (SS) spectral type, while all early-red events Ðexcept for the 2000cx-like SN~2012fr are of the Branch Core-Normal (CN) or CooL (CL) type. 
A number of potential processes contributing to the early emission are explored, and we find that, in general, the viewing-angle dependance inherent in the companion collision model is inconsistent with all SS objects with early-time observations being blue and exhibiting an excess.  We caution that great care must be taken when interpreting early-phase light curves as there may be a variety of  physical processes that are possibly at play and significant theoretical work remains to be done.
\end{abstract}


\section{Introduction} \label{sec:intro}

Type~Ia supernovae (SNe~Ia) are well-studied astrophysical events generally thought to be the thermonuclear disruption of a carbon-oxygen (C/O) white-dwarf (WD) in a binary system.  
In order for future SN~Ia experiments to  expand upon our knowledge of dark energy, a substantial improvement in their accuracy as distance indicators is required.
This is  likely only to be achieved by increasing our understanding of SNe~Ia progenitors and their explosion physics. 
Early-phase observations of SNe~Ia offer a unique window to better understand their origins.  To date early observations have allowed for a direct constraint on the size of the WD progenitor of SN~2011fe \citep{nugent11,bloom12} using the lack of a shock cooling signal \citep{piro10}, and in about a dozen cases, have provided robust  constraints on the size of any potential companion \citep{foley12,bloom12,silverman12,zheng13,goobar15,olling15,im15,marion16,shappee16,cartier17,hosseinzadeh17,holmbo18,miller18,shappee18a}.  

\begin{figure*}[ht]
\begin{center}
\includegraphics[width=12.cm,angle=270]{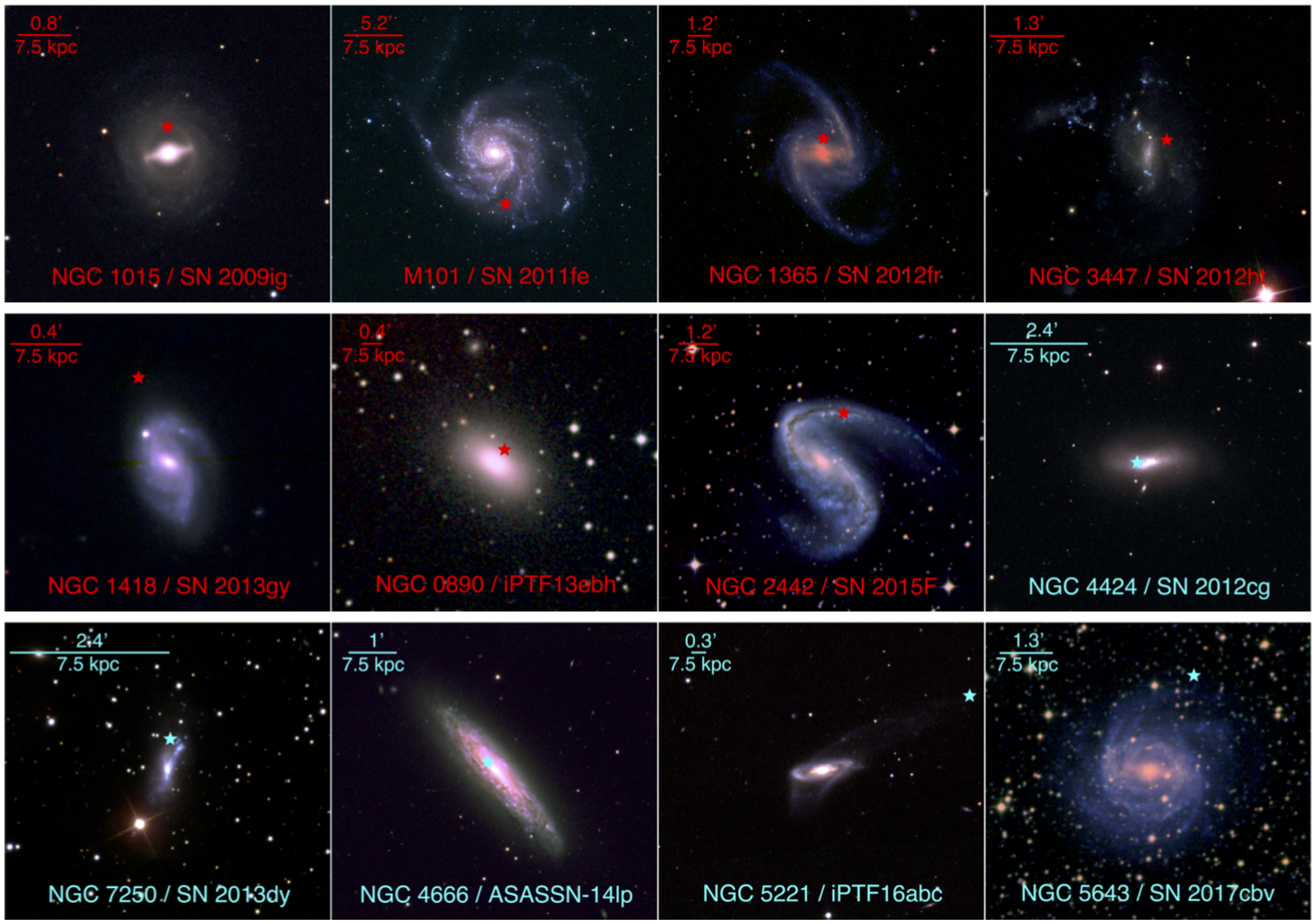}
\caption{A mosaic of colored images of our early red vs. blue  sample  with the position of the SNe indicated with stars. Images were constructed using either  SDSS $gri$-band, DSS  IR-red-blue, or (in one case) Pan-STARRS $gri$-band archival images. \label{fig:FCs}}
\end{center}
\end{figure*}


 To date nearly 20 SNe~Ia have been discovered within three days of their inferred time of first light ($t_{first}$)\footnote{In this work we refer to $t_{first}$ as the explosion epoch, however, it is possibly that some SN~Ia  experience a 1-2 day dark phase between the explosion and  $t_{first}$ \citep{piro13}.},  and this sample exhibits interesting diversity. The early light curves of one group rise exponentially and are typically well-fit by a single power law function \citep[e.g.,][]{nugent11,olling15}. In a second group, the early light curves exhibit a  $\approx$ 3 day linear rise in flux followed by an exponential rise \citep[e.g.,][]{hosseinzadeh17,miller18,contreras18}. Such objects are well fit with a double (or broken) power law fit \citep[e.g.,][]{zheng13,zheng14}. It is a matter of open debate what physics is driving single vs. double power-law fit SN~Ia ,with possibilities ranging from  companion interaction  \citep{kasen10,maeda14} to enhanced mixing of  radioactive elements  \citep{piro13,piro16,magee18}. 
Furthermore, the full diversity of early phase properties is likely far from fully explored.  This point is highlighted by the early light curve of MUSSES1604.  In this case the supernova exhibited an optical flash associated with a $m_g$ $\approx 2$ mag increase in brightness within 24 hours of explosion, followed by a short, 24 hour plateau period, and then subsequently, it continued to brighten similarly to other SNe~Ia \citep{jiang17}. 

In the following, we collect a sample of early phase SN~Ia observations in order to examine their intrinsic $(B-V)_{0}$ color evolution. We then examine the location of the sample objects along the Phillips relation  and the \citet{branch06} diagram. We find that SNe~Ia exhibiting blue, slowly evolving  $(B-V)_{0}$ colors are also of the Branch shallow silicon spectral type as  compared to  red, more rapidly evolving objects. This suggests at least two distinct populations of SNe Ia. The host properties of the sample are also examined and our findings are placed into context with leading models.

\vspace{1.0cm}
\section{Data sample}

We have combed the literature to identify all objects useful to explore the diversity among the early color evolution of  SNe~Ia. 
Two selection criteria were imposed. First the discovery must have occurred within three days of $t_{first}$, and secondly, each object has early $B$- and  $V$-band light curves.
We found 13 objects fulfilling these  criteria and present them in Table~\ref{tab:sampleparameters}, along with a number of pertinent details  including:  host designation and redshift, Milky Way and host-galaxy reddening, inferred $t_{first}$, and $t_{rise}$.  Additionally, estimates of the light curve decline-rate parameter $\Delta m_{15}(B)$\footnote{This parameter is the difference in magnitude between $B$-band peak and 15 days later and is known to correlate with $M_B$ \citep{phillips93}.}; peak absolute $B$-band magnitude ($M_B$); \citet{branch06}  and \citet{wang09} spectral types;  our early-color sub-typing (i.e., red or blue; see \S~\ref{sec:color}); and references to adopted host-reddening values, data and/or adopted values of $t_{first}$. 
\begin{figure}[ht]
  \includegraphics[width=0.5\textwidth]{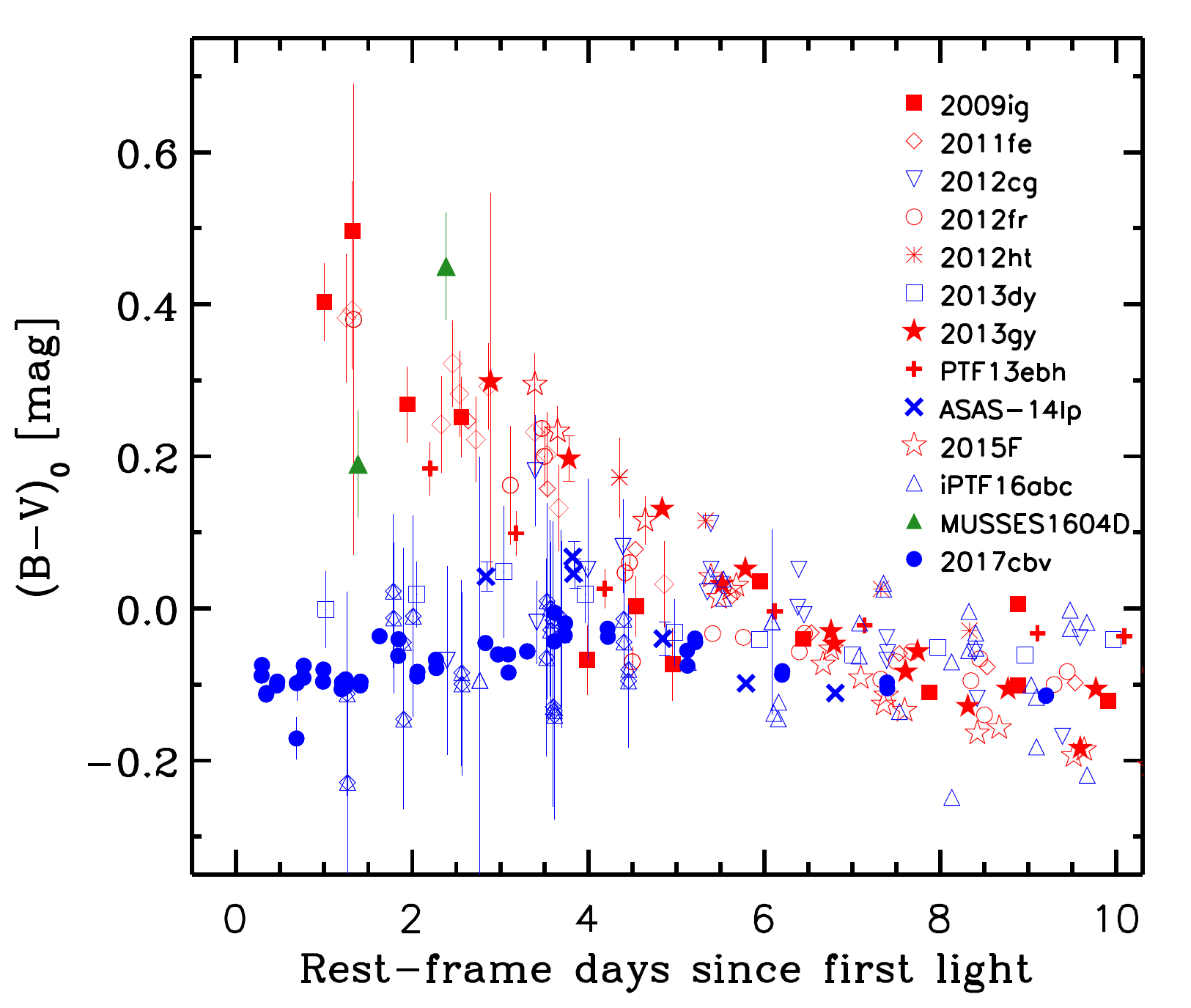}
  \caption{Optical $(B-V)_{0}$ color evolution for SNe~Ia discovered within 3 days of first light, $t_{first}$. For presentation purposes error bars are included for all colors obtained within 5 days of $t_{first}$. The early evolution of the current sample indicates SNe~Ia exhibit either red colors that rapidly evolving to the blue, or blue colors that  evolve relatively slow over time.
Time-dilation and reddening corrections have been applied using red-shift and extinction values listed in Table~\ref{tab:sampleparameters}.}
\label{fig:color}
\end{figure}

We note that two members of the early sample --SN 2009ig and SN 2012fr--    have been pointed out to be somewhat peculiar  \citep{contreras18} as they exhibit  similarities  to SN~2000cx \citep{li01}. 
These two objects are include in   our early sample since both objects are fully consistent with the luminosity decline rate relation \citep[see][and below]{foley12,contreras18} and their early light curve evolution is not peculiar as in the case of MUSSES1604D.
MUSSES1604D itself is a 2006bt-like event \citep[e.g.,][]{foley10,stritzinger11} with  photometric and spectroscopic characteristics that differ significantly at early and maximum phase compared to the rest of our sample \citep{jiang17}.  
However, as it does fulfil our selection criteria it is  included in our sample for completeness as a green symbol in the figures presented below. 

Figure~\ref{fig:FCs} contains a color image of each host galaxy of the early sample  with the location of the supernova indicated with a star (color coded red or blue; see below).   According to NED, all members of our early sample are hosted in spiral galaxies, except the host of MUSSES1604D, which is an S0 galaxy \citep{jiang17}.
No obvious trends are found between early color type and host properties and/or the locations of the supernovae relative to their hosts.

\section{Results}
\subsection{Early sample $(B-V)_0$ color evolution}
\label{sec:color}

Figure~\ref{fig:color} contains the intrinsic $(B-V)_0$ color evolution for the early sample, color-coded in either red or blue. The color curves have been corrected for both Milky Way reddening, host-galaxy reddening and time dilation.
Red objects exhibit $(B-V)_{0}$ $\gtrsim 0.2$ mag by $+$2~d, and as much as $\approx$0.5 mag at $+$0.5~d. The color of the blue objects  ranges between $(B-V)_0~\sim~-0.2$ to 0.05 mag within the first $+$2~d relative to $t_{first}$ and evolve relatively slowly compared to the red objects. By $+$4~d the color difference between the two groups is negligible.

Inspection of the color difference between the two populations reveal differences on the order of $\sim$0.5 mag, which corresponds to a difference in flux of $\sim$50\%. This is   far too large of a flux difference to be attributed to spectral line features.  Due to a scarcity of data, here we focus exclusively on  $(B-V)_{0}$, however, we  note that the blue vs red sub-types holds over different color combinations \citep[see, e.g.,][]{hosseinzadeh17}.

\begin{figure}[ht]
\includegraphics[width=0.5\textwidth]{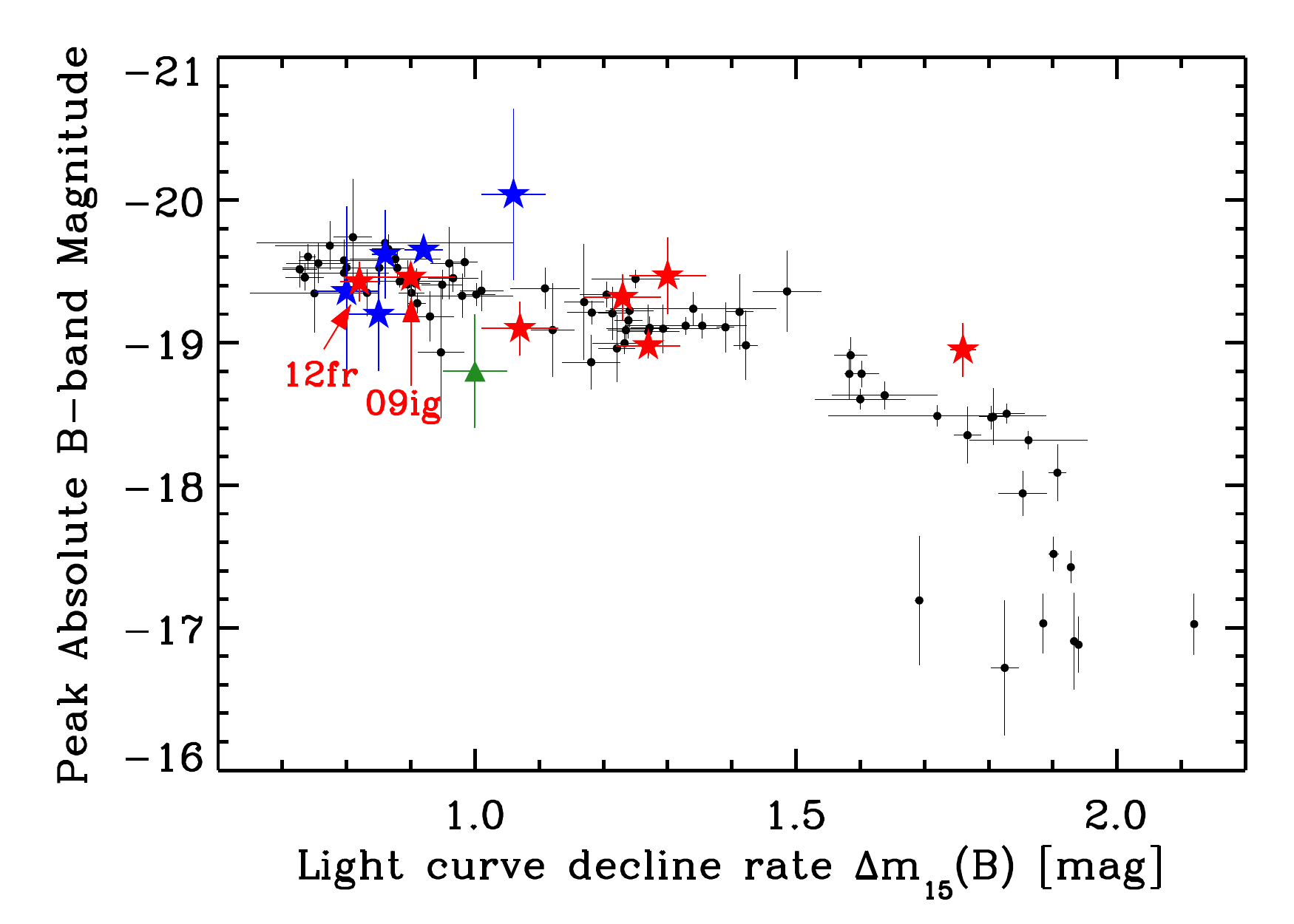}
\caption{Luminosity decline-rate relation consisting of 1991T-like, normal and 1991bg-like SNe~Ia observed by the Carnegie Supernova Project-I \citep{krisciunas17}. Color stars correspond to the  comparison sample with blue stars corresponding to objects with blue colors at early phases and red stars are those exhibiting red colors.   Excluding the peculiar 2000cx-like SNe~2009ig and 2012fr,  objects with early, slowly evolving blue colors are generally brighter and exhibit slower declining light curves than  objects  with early, rapidly evolving red colors.  }
\label{fig:dm15}
\end{figure}

\subsection{Early sample on the $M_B$ vs. $\Delta m_{15}(B)$ relation}

Plotted in Figure~\ref{fig:dm15} are the $M_B$ and $\Delta m_{15}(B)$ values listed in Table~\ref{tab:sampleparameters} of the early sample along with  a sample of  1991T-like, normal, and 1991bg-like  supernovae observed by the Carnegie Supernova Project I \citep{krisciunas17}. Objects exhibiting early blue colors tend to exhibit higher peak luminosities and slower decline rates. Interestingly, the two red objects   that are  as bright as the majority of the early blue objects are both 2000cx-like objects \citep{contreras18}.

\subsection{Early sample on the Branch diagram}

We now examine   the ratio of the pseudo-Equivalent Widths  (pEWs) measured for the \ion{Si}{2}   $\lambda$5972 and \ion{Si}{2} $\lambda$6355 spectral features in  optical spectra  obtained within three days of $B$-band maximum. This ratio is known to correlate with photospheric temperature \citep{nugent95} and serves as basis for the  \citet{branch06} spectral classifications consisting of:  Core Normal (CN), Broad Lined (BL), Shallow Silicon (SS), and CooL (CL). CN are normal  SN~Ia. BL show higher than normal \ion{Si}{2} $\lambda$6355 Doppler velocity. 
Among the BL objects, two-thirds correspond to high-velocity (HV) objects in the \citet{wang09} classification scheme, as defined by exhibiting  \ion{Si}{2} $\lambda$6355  Doppler velocity of $\geq$10,800 km~s$^{-1}$ at maximum light.
Finally, SS are typically bright 1991T/1999aa-like objects and the  CL sub-type contains both transitional  and 1991bg-like SN~Ia. 

\begin{figure}[h]
\includegraphics[width=0.5\textwidth]{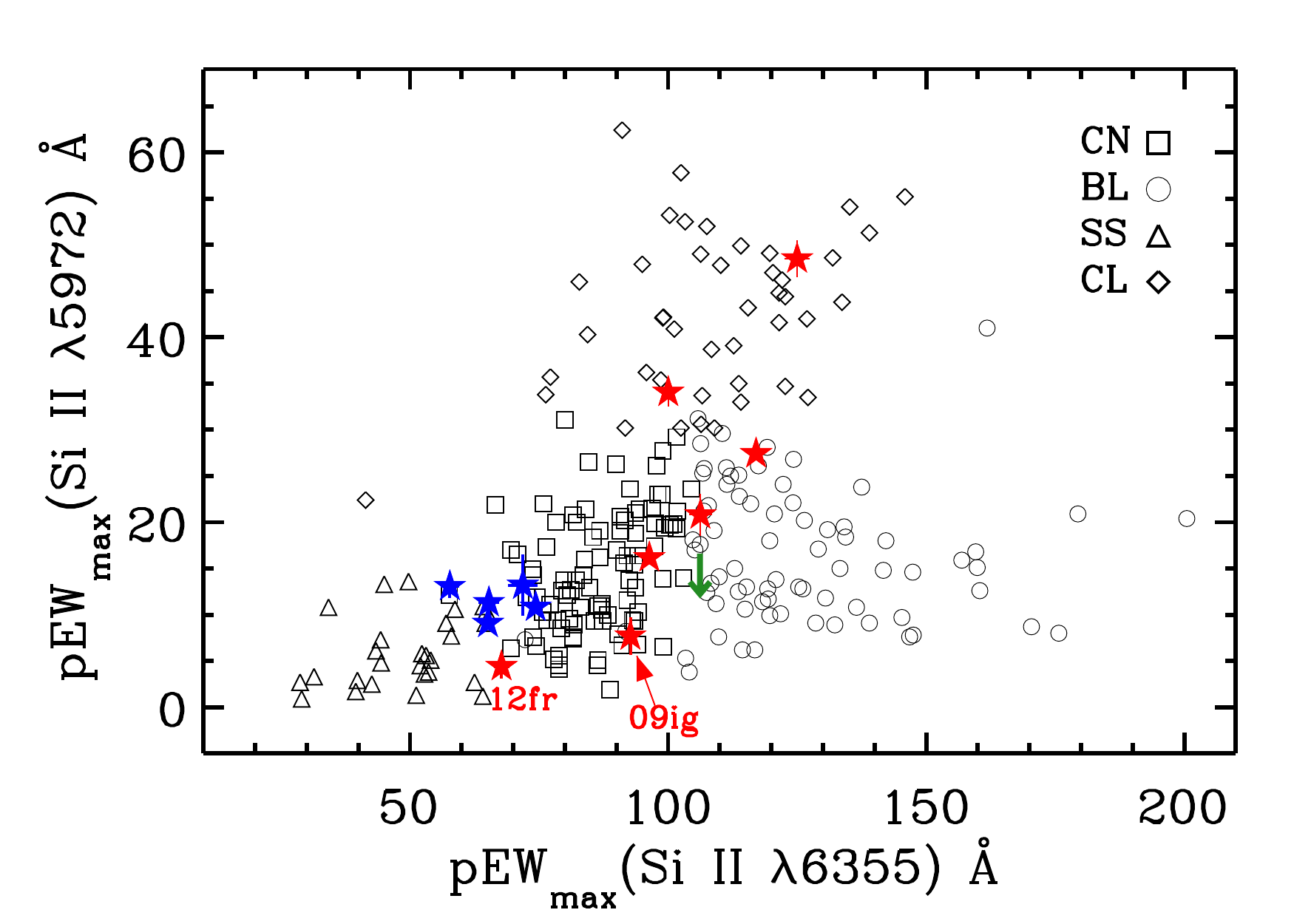}
\caption{Branch diagram of the pEW values measured from the  \ion{Si}{2} $\lambda$5972  and  \ion{Si}{2} $\lambda$6355 absorption features in near maximum light spectra from the  \citet{blondin12} sample along with pEW values measured of our early sample.  The four Branch spectral types are indicated with different black symbols and the early sample  is plotted as colored stars. All normal SNe~Ia exhibiting blue colors are either of the shallow silicon (SS) sub-type or lie just along the interface between the SS and CN (core normal) types. Note  the $\lambda$5972 pEW measurement for MUSSES1604D is an upper limit.}  
\label{fig:branch}
\end{figure}

Figure~\ref{fig:branch} contains the Branch diagram populated with an extended sample of SNe~Ia \citep{blondin12}, along with our early sample color-coded red vs. blue. 
Interestingly, aside from the peculiar 2000cx-like SN~2012fr \citep{contreras18}, all of the red objects are of the CN or CL sub-type, and all of the blue objects are located in a relatively narrow  region extending from SS to the interface with the edge of the CN population. 

Although SN~2012fr follows the luminosity decline-rate relation (see Fig.~\ref{fig:dm15}), it is spectroscopically peculiar. 
At maximum light its \ion{Si}{2} $\lambda$6355 Doppler velocity  is 12,000~km~s$^{-1}$ \citep{childress13}. This places SN~2012fr as a HV object in the  \citeauthor{wang09} system,  while its \ion{Si}{2} velocity remains essentially flat for a month past maximum  \citep{childress13}. Following \citet{benetti05}, SN~2012fr is a low-velocity gradient (LVG, i.e., velocity gradient $\dot v <  70$ km s$^{-1}$ day$^{-1}$) object.  Only  $\approx 10\%$ of SNe~Ia are found to be  HV  and LVG, while  only $\approx$ 5\% of SNe~Ia are also HV and SS+CN  \citep[see][]{contreras18}. Altogether with its  peculiar spectroscopic attributes,  we note one should approach SN~2012fr with caution when comparing its spectral diagnostics to our limited early sample.

\section{Discussion}

 Using the early phase light curve colors, we have identified at least two distinct populations of SNe Ia, the properties of which are as follows:
\begin{itemize}
\item Events that are blue at early times have spectra similar to 1991T/1999aa-like objects, fall within the Branch SS spectral type, and are typically more luminous at peak with a smaller $\Delta m_{15}(B)$.

\item Events that are red at early times are mostly of the Branch CN or CL spectral types (with the exception of one 2000cx-like object) and may typically be less luminous at peak with a larger $\Delta m_{15}(B)$.
\end{itemize}

We now briefly discuss various processes that may be contributing to the early phase emission. These range from interaction with a non-degenerate companion, the presence of high-velocity $^{56}$Ni, interaction with circumstellar material (CSM), and opacity differences in the outer layers of the ejecta.

\subsection{Interaction with a non-degenerate companion}

 \citet{kasen10} presented simulations of an SN~Ia interacting with a non-degenerate companion. This creates a shock cooling signature over the first few days, and has been a popular explanation for many of the early-blue events \citep[e.g.,][]{marion16,hosseinzadeh17}.
 Other similar simulations predict a lower luminosity and shorter duration \citep{marietta00,maeda14,kutsuna15}, but are still potentially consistent with the early-blue events depending on the size and distance of the companion.
 
 If interaction with a non-degenerate companion produces early blue colors then one is left to explain why these objects have a preference to be 1991T/1999aa-like and of the Branch SS type.
Moreover, the interaction model predicts the blue excess to be strongest when the companion sits between the explosion and the observer, so that less favorable viewing angles should still look red. Thus if this is the preferred explanation, then there should be early red events that are also bright and of the Branch SS type. However, such objects have yet to be discovered.
 
 \subsection{Presence of high-velocity $^{56}$Ni}

 Another way to produce additional heating at early times is if there is $^{56}$Ni located  in the very outer layers. This could be due to significant mixing by instabilities or large scale flows during the explosive burning. Alternatively, $^{56}$Ni can be deposited in the outer layers from a helium-shell triggered double-detonation explosion \citep{woosley80,nomoto82}, which can occur when there is a helium donor or even due to the helium-rich surface layers of a C/O WD during a double degenerate merger. Such scenarios have been explored by \citet{piro16} and \citet{magee18}, which find similar timescales, luminosities, and colors to the early-blue events when there is $\sim0.01\,M_\odot$ of $^{56}$Ni in the surface layers.

The main criticism of the shallow $^{56}$Ni explanation is whether a large abundance of iron group elements (IGEs) in the surface layers negatively impacts the spectra of SNe Ia at peak \citep{hoeflich96,nugent97}. More detailed treatments of the helium detonation  \citep{shen14,shen18} and the converging shock in the C/O core \citep{shen14b} find that this scenario may still be viable with smaller helium masses and less IGEs, but additional work is needed to understand whether  it can both explain the early blue phase and still be consistent with the full spectral evolution of the events. Furthermore, the current set of observations and models  do not have the fidelity  to distinguish between a smooth, shallow mixed distribution of $^{56}$Ni as focused on by \citet{piro16} and \citet{magee18} or a distinct blob of $^{56}$Ni located near the surface as might be expected for  a double-detonation \citep[see][]{maeda18}.

A related but slightly different way to produce additional early energy release is if there is nuclear burning in the outer layers \citep{hoeflich09}. This differs from the double-detonation scenario in that a thin helium layer of $\sim10^{-5}M_{\sun}$ mixed with some carbon \citep{wang17} is triggered by the detonation front propagating out of the C/O core (rather than the helium layer igniting first). It has also been speculated that the helium layer could be triggered by {\it g}-modes driven by the convective simmering phase of carbon burning in single-degenerate scenarios \citep{piro11}. In contrast, a hydrogen-rich layer would not burn due to  the longer burning time scales which are well in excess of 1~s, compared to 10$^{-2}\,$s for  a mixed helium-carbon layer.

\subsection{Circumstellar Dust}
Another potential source that could affect the early colors is varying amounts of circumstellar dust \citep{amanullah11}. 
 However, this would require the dust to be fairly isotropic to avoid the unlikely proposition that the dust in the direction of the observer is always the same. Furthermore, the dust cannot impact the colors at peak which would require small patches of consistently well-aligned dust.  While circumstellar dust may play a role in increasing the scatter in the color of early-time light curves we do not believe it can create a correlation of early color with other observational properties as presented in this work.

\subsection{Circumstellar interaction}

Yet another method for producing additional heat at early times is through the interaction of the ejecta with CSM. This could occur in the collision of the ejecta with distinct shells of material that are from the companion or pulsational events during the explosion itself. Taking a typical expansion velocity in the corresponding outer layers ($\sim$40,000 km~s$^{-1}$)  equates to a distance of $\sim$10$^{15}\,{\rm cm}$ \citep{hoeflich02} within 3-5 days of $t_{first}$. With this scale in mind, early emission could be produced from the conversion of kinetic energy of the ejecta into heating  through its interaction with CSM \citep{gerardy07, dragulin16,noebauer16}.  However, due to the short time scales covered by our early sample this is likely not a viable process as  the CSM  would have to be bound to the system of size less than $\sim10^{15}$cm, or in the direct vicinity.

Alternatively, the CSM could be more confined to the WD such as the tidally disrupted material that is present after a double-degenerate merger or the accretion flows in a single-degenerate scenario. In such cases, if the material is optically thick to the explosive shock wave, then the shock continues to propagate into the CSM and heat it \citep{piro16}. This produces a shock cooling signature at early times, with a luminosity that is proportional to the radius of the CSM. For the typical early luminosities of the early-blue events, this implies a radius of $\sim10^{11}-10^{12}\,{\rm cm}$ for such material. Further work is needed to better understand if the CSM would impact the spectra and light curve evolution in other ways that can be tested with observations.

\subsection{Composition/opacity differences}

A final possibility we consider is whether the early blue excess could actually not be due to additional energy input but instead simply differences in opacity due to a different composition.  If the outer  layer of the ejecta is characterized by a lower opacity (e.g., due to a significant amount of unburnt carbon), the  photosphere would recede more rapidly and thus   reach  the $^{56}$Ni-rich region earlier. The overall affect of this is a faster release of energy, more heating, and thus bluer colors. This affect is demonstrated in 
\citet[][see their Figure~11]{gall18}, where  $10^{-2}\,M_{\sun}$  of unburnt carbon  produces an earlier UV-blue phase lasting  $\approx$4 days.

\subsection{Closing remarks}

In conclusion,  from the perspective of early phase colors, there appears to be at least two populations of SN~Ia. Looking toward the future, the key question will be whether this represents a critical difference between the progenitors of these events, or if it is the result of smaller differences in composition, explosion energy, or some other detail of the explosions. Solving this mystery will require further theoretical modeling as well as gathering a larger sample of events to see how strongly this dichotomy persists and whether there are other properties that correlate with the early red and blue populations.

\textit{While this manuscript was under review \citet{jiang18} presented an examination of  early-excess  SNe~Ia. They also find a  preference  of blue, early-excess SNe~Ia to be 1991T/1999aa-like.   Their interpretation is consistent and complementary to our findings.}

\begin{deluxetable*}{lccccccccccc}
\tabletypesize{\tiny}
\tablewidth{0pt}
\tablecaption{Early Color Evolution Sample Parameters.\label{tab:sampleparameters}}
\tablehead{
\colhead{SN} &
\colhead{Host} &
 \colhead{Red-shift} &
\colhead{$E(B-V)_{MW}$} &
\colhead{$E(B-V)_{host}$} &
\colhead{$t_{first}$} &
\colhead{$t_{rise}$} &
\colhead{$\Delta$m$_{15}(B)$ } &
\colhead{$M_B$ } &
\colhead{Spectral-type\tablenotemark{a}} &
\colhead{Color} &
\colhead{References(s)} \\
\colhead{} &
\colhead{} &
\colhead{} &
\colhead{[mag]} &
\colhead{[mag] } &
\colhead{[MJD]} &
\colhead{[days]} &
\colhead{[mag]} &
\colhead{[mag]} &
\colhead{} &
\colhead{}     &
\colhead{} }
\startdata
2009ig     & NGC 1015 & 0.00877 & 0.032 & \nodata    &  55062.9 & 17.1 & $0.90\pm0.07$ & $-19.46\pm0.12$  & CN, HV      & red & (1) \\ 
2011fe     & M101     & 0.00080 & 0.008 & \nodata&  55796.7 & 17.8 & $1.07\pm0.06$  & $-19.10\pm0.19$ & CN, normal  & red & (2)  \\ 
2012cg     & NGC 4424 & 0.00146 & 0.018 & $0.18\pm0.05$\tablenotemark{b} &  56061.8 & 19.5 & $0.86\pm0.02$ & $-19.62\pm0.31$  & SS, 91T/99aa-like  & blue & (3) \\ 
2012fr     & NGC 1365 & 0.00546 & 0.018 & $0.03\pm0.04$\tablenotemark{c} &  56225.8 & 16.9 & $0.82\pm0.03$ & $-19.43\pm 0.14$  & CN+SS, HV   & red  & (4) \\ 
2012ht     & NGC 3447 & 0.00356 & 0.025 & $0.05\pm0.01$\tablenotemark{c} &  56278.0 & 17.6 & $1.27\pm0.05$ & $-18.98\pm0.07$  & CN, normal          & red  &  (5)   \\
2013dy     & NGC 7250 & 0.00389 & 0.140 & $0.21\pm0.01$\tablenotemark{d} &  56483.4 & 17.7 & $0.92\pm0.03$ & $-19.65\pm0.04$  & SS, 91T/99aa-like  & blue & (6) \\ 
2013gy     & NGC~1418 & 0.01402 & 0.049 & $0.11\pm0.06$\tablenotemark{c} &  56629.4 & 19.1 & $1.23\pm0.06$ & $-19.32\pm0.16$ & CN, normal  & red  & (7)  \\ 
iPTF~13ebh & NGC 890  & 0.01327 & 0.067 & $0.07\pm0.02$\tablenotemark{c} & 56607.9  & 14.8 & $1.76\pm0.02$ & $-18.95\pm0.19$  & CL, normal   & red  & (8)  \\ 
ASASSN-14lp& NGC 4666 & 0.00510 & 0.021 & $0.35\pm0.01$\tablenotemark{c} &  56998.5 & 16.7 & $0.80\pm0.05$ & $-19.36\pm 0.60$  & SS, 91T-like  & blue & (9) \\ 
2015F      & NGC 2442 & 0.00489 & 0.175 & $0.16\pm0.03$\tablenotemark{c} &  57088.4 & 18.5 & $1.25\pm0.05$ & $-19.47\pm0.27$  & CN, normal  & red  & (10) \\ 
iPTF16abc  & NGC 5221 & 0.02328 & 0.028 & $0.05\pm0.03$\tablenotemark{e} &  57481.6 & 17.9 & $0.85\pm0.05$ & $-19.20\pm0.40$  & SS, normal/91T-like & blue & (11) \\ 
MUSSES1604D & \nodata & 0.11737 & 0.026  & \nodata      &  57481.8 & 22.4 & $1.00\pm0.07$ & $-18.80\pm0.40$   & BL, HV  & red & (12) \\ 
2017cbv    & NGC 5643 & 0.00400 & 0.150  & \nodata      &  57821.9 & 18.3 & $1.06\pm0.05$ & $-20.04\pm0.60$   & SS, 91T-like  & blue & (13)  \\ 
\enddata
\tablenotetext{(a)}{~~\citet{branch06} and \citet{wang09} spectral classifications.}
\tablenotetext{(b)}{~~\citet{silverman12}.}
\tablenotetext{(c)}{~~Based on SNooPy fits to photometry obtained by the Carnegie Supernova Project-II (Burns et al., in prep).}
\tablenotetext{(d)}{~~\citet{pan15}.}
\tablenotetext{(e)}{~~\citet{miller18}.}
\tablecomments{(1) \citet{blondin12}, \citet{foley12}, \citet{marion13}; (2) \citet{nugent11}, \citet{vinko12}, \citet{pereira13}; (3) \citet{silverman12}, \citet{marion16}; (4) \citet{contreras18}; (5) \citet{yamanaka14}, This work;  (6) \citet{zheng13}, \citet{pan15}, \citet{zhai16}; (7) \citet{holmbo18}; (8) \citet{hsiao15}, (9) \citet{shappee16}, This work; (10) \citet{cartier17}, This work; (11) \citet{miller18}, This work; (12) \citep{jiang17}, This work; (13) \citet{hosseinzadeh17}, This work.}
\end{deluxetable*}

\acknowledgments
We thank P. Brown, G. Hosseinzadeh, J. Jiang,  and A. Miller  for promptly providing access to published data of several of the objects in our early sample.
This work has been supported by a research grant 13261 (PI Stritzinger) from the Villum FONDEN. M. D. Stritzinger is grateful to Aarhus University's Faculty of Science \& Technology for a generous sabbatical grant. C. Burns, N. Morrell, A. Piro, and M. Phillips  acknowledge support from the National Science Foundation under grant AST1613426.
E. Baron and P. Hoeflich gratefully acknowledge support from NASA Grant NNX16AB25G. P. Hoeflich also acknowledge the NSF grants AST-1715133 and AST-1613472.



\end{document}